\documentclass[12pt]{article}
\usepackage{emlines2,bezier}
\usepackage{amssymb}
\author{\normalsize Sergey S. Kokarev\thanks{email: sergey@yspu.yar.ru}\\
\small\it Department of theoretical physics, r.409, Yaroslavl State Pedagogical
University,\\ \small\it Respublikanskaya 108, Yaroslavl, 150000, Russia}
\title{\large\bf Classical solids dynamics as 4D statics of elastic strings}
\date{}
\def\bm#1{\mbox{\boldmath$#1$}}

\begin{document}
\maketitle \baselineskip=18pt

\begin{abstract}\small
Variational principle for a solid in classical mechanics is formulated in terms
of a thin elastic 4D bar strain  in events space $\mathbb{M}_4$ of special
relativity. It is shown, that the sum of elastic 4-energies of weak twist and
bending under some identifications takes the form of classical non-relativistic
action for a solids dynamics. The necessary conditions on 4D bar parameters and
elastic constants, providing validity  of Newton mechanics, are found.
\end{abstract}

\vspace{0.8cm} {\small PACS: 46.70.Hg; 46.25.Cc; 45.20.Jj; 03.30.+p.}

\section{INTRODUCTION}

In recent years  extended  objects with small effective dimensions --- strings
and membranes --- hold central position in modern theoretical phy\-sics
\cite{green}-\cite{rand}. Due to a rich physical and geometrical content, they
are, probably, the most suitable candidates for unified models of space-time,
particles and fields. Apart from the unification ideas (or partly due to ones),
the string  and brane concepts suggest that there are many physical topics,
considered earlier as quite different, which have deep interrelations between
each other. An example is intriguing  physical analogy between elasticity
theory \cite{land1} and Einstein gravity \cite{land2}, originally noted by Born
\cite{born}. Based on this analogy new approach to a space-time-matter dynamics
has been developed in works \cite{kok1,kok2,kok3,kok4}. It has required some
reformulations  of the both theories: elasticity theory has been generalized to
multidimensional elastic bodies and Einstein gravity (Einstein-Gilbert action
for gravitational field) has been written in terms of isometric embedding
\cite{eiz}. Investigation of the analogy within the context of the approach
reveals very simple physical interpreting of GR action: {\it total action
\begin{equation}\label{act}
\mathfrak{s}=\mathfrak{s}_{\rm g}+\mathfrak{s}_{\rm m}
\end{equation}
for a system  "gravitational field + matter" has  mechanical sense of
multidimensional elastic energy of a strained thin 4D plate:
\begin{equation}\label{frengr}
\mathfrak{F}=\mathfrak{F}_{\rm b}+\mathfrak{F}_{\rm s},
\end{equation}
where the first term  is responsible for its bending, second --- for tangent to
its surface stretches-shears.} The term "4D plate"\footnote{Here and below we
distinguish strongly tensed objects (strings and membranes) from a more general
(bars and plates) arbitrarily stressed ones.} means multidimensional body,
whose sizes along some 4D Riemannian manifold $\mathbb{V}_4$ are much more than
orthogonal to $\mathbb{V}_4$ ones. Status analysis of a number of objects
within  classical field theory (lagrangian, metrics and metric energy-momentum
ten\-sor, spa\-ce-ti\-me signature, boundary conditions) has been made in frame
of the approach in  \cite{kok3}. Dimensional analysis gives the following
relation between Einstein constant $\varkappa$  and elastic constants of the
plate
--- Young modulus and thicknesses:
$$ \frac{1}{\varkappa}\sim Eh^{N+3}, $$ where  $N$ --- number of
extradimensions. It supports old Sacharov's idea about elastic origin of
Einstein constant $\varkappa$ \cite{sah}. Embedding theory and elasticity
theory have been applied earlier independently in a number of works on gravity,
brane physics and quantum theory \cite{pav3}  (and ref-s therein),
\cite{tar,hehl,dmitr}.

If the discussed analogy is not occasional, then multidimensional elasticity
should be relevant also to clearing of physical picture in case of simpler
systems. In present article we apply the approach to solids dynamics in
classical mechanics. In other words, our aim is derivation of mechanical action
from the multidimensional elastic energy functional.  Our consideration will
naturally lead us again to the 4D objects with small (equal to 1) effective
dimension --- {\it thin bars}, posed in  4D Minkowski space, which from now on
will be shortly referred to  as {\it 4-bars.} These objects are of different
nature, than ordinary classical strings, considered within the field theory.
Following analysis  shows, that the 4-bars deformation theory reproduces
classical Newton mechanics, when the bars posses a very strong tension. By
standard terminology of elasticity theory, we call such strongly tensed bars
{\it strings.} Its difference from the ordinary ones is that they are time-like
and correspond to world tubes of finite space-like thickness within special
relativity (SR), while classical string is a space-like 1D object (i.e. has
nonzero length and zero thickness)\footnote{Though moving ordinary string
generates two-dimensional matter sheet in $\mathbb{M}_4$, it possess  a null
thickness and so should be related to a more particular subclass of a bars:
tubes (closed string) or bands (open string).}. On the other hand, in contrast
to  world tubes of SR, the strings studied  here possess 4D elasticity, in
particular, time-like one. To distinguish such strings from standard ones,
we'll shortly call them through the paper {\it 4-strings,} reminding that we
deal with 4D object. Note, that our approach overlaps with Pav\u{s}i\u{c} works
\cite{pav1,pav2} in two points: we use the similar concept of observer
(simultaneity hypersurface and its motion in $\mathbb{M}_4$) and get observable
3-D classical equations of motions as consequence of a more general
multidimensional physics with "frozen" matter.

The paper is organized as follows. In Sec.\ref{rel} we give sketch of events
space properties and 4D kinematics within SR. Here we also introduce the
definition of general 4-body by analogy with 3D case.

Sec.\ref{thbar} restricts our attention to the important subclass of 4-bodies
--- thin 4-bars, which are 4D ``pre-images'' of observable 3-bodies in
classical mechanics.

Sec.\ref{4strain} is devoted to a 4D generalization of standard elasticity
theory.

In Sec.\ref{torsion} we derive  elastic 4-energy functional for the case of
pure twist of a thin 4-bar. We show, that the 4-energy goes into classical
rotational  part of mechanical action, when 4-bars space sections remain
unstrained (rigid 3-body) and the twist is weak (non-relativistic rotation).

Weak bending of a 4-bar is investigated in Sec.\ref{bend}. We give general
relativistic description of 4-bar stretchless strain, including both bending
and twist and then extract the part, which  is responsible for pure bending. It
turns out, that the elastic energy of bending is quadratically depends on
acceleration of 3-body, so the equation of motion  of 3-bodies  (static
equilibrium of 4D ones) should have a forth order. To reject this contradiction
to experience, we go from 4-bars to 4-strings  and require negligibility of its
bending energy in comparison with stretching one.  It gives us additional
restriction, clearing 4-string physics (see noneq.(\ref{est}) --- strong
tension condition) and right action of classical mechanics --- quadratic over
velocities in nonrelativistic case.

Sec.\ref{concl} contains qualitative discussion of classical mechanics within
the frame of developed approach.

Throughout the whole paper, unless otherwise specified, standard indexless
notations are used. Particularly, small Latin, Greek or Gothic letters denote
3D objects, capital ones --- corresponding 4D ones.

\section{SOLIDS KINEMATICS IN SPECIAL RELATIVITY}\label{rel}

The aim of this section is to remind  mathematical properties  of  events space
of SR, which lies in  foundation of 4D bar mechanics. SR endows affine space of
events $\mathbb{A}_4$ by the {\it Lorentzian structure}, which contains {\it
pseudoeuclidian structure} and {\it relativistic symmetry group}. The first
defines distance $S_{AB}$ between any two points-events $A,B\in\mathbb{A}_4$:
$$ S_{AB}=S_{BA}=\sqrt{({\bm V}_{AB},{\bm V}_{AB})},$$ where  ${\bm V}_{AB}$
--- 4-vector, (element of affine vector space $\mathbb{W}_4)$ and scalar
product is determined by {\it Minkowski metrics} with signature $(+,-,-,-)$.
The hyperbolicity of  Minkowski metric leads to the three type of vectors in
$\mathbb{W}_4$: {\it time-like} ($({\bm V},{\bm V})>0$), {\it space-like}
($({\bm V},{\bm V})<0$) and {\it iso\-tro\-pic} $(({\bm V},{\bm V})=0$).

The second --- {\it Poincare group} $\mathrm{P}$ ---  is subgroup  of a group
of general non-homogeneous linear transformations  $\mathrm{GL(4,R)}$ in
$\mathbb{A}_4$, which don't change Minkowski metrics. Space of events with
Minkowski metric and isometry Poincare group is called {\it space-time}
$\mathbb{M}_4$.

Third element, we need to introduce, is related to an  observable  in
$\mathbb{M}_4 $ physics. Let $A,B\in\mathbb{M}_4$ --- pair of events, connected
by a simple smooth path $\gamma_{AB}$. Then time interval  $T_{AB}$ between
$A$  and $B$ along $\gamma_{AB}$ is the integral
$$\int\limits_{\gamma_{AB}}\tilde{\bm T},$$ where $\tilde{\bm  T}=\tilde {\bm
T}_{\mu}dX^{\mu}$
--- {\it relativistic form of time}, also referred to in papers on the subject as
$\tau$-field  or reference frame form \cite{mitsk}. In a difference with
non-relativistic mechanics, where form  $\tilde{\bm  t}$ is unique (absolute)
for all reference frame,  number forms of time in  $\mathbb{M}_4$ is infinitely
large. Note, that absence of the form $\tilde{\bm{T}}$ within the SR,  makes it
physically contentless, since we lose any possibility to compare the theory
with experiment\footnote{So we follow to the Arnold's book \cite{arnold}, where
similar three blocks of classical Newton mechanics --- Galilean structure,
Galilean symmetry group and form of time --- are accented.}. The only algebraic
condition   imposed on $\tilde{\bm  T}$ is $$ (\tilde{\bm  T},\tilde{\bm
T})=1, $$ which means, that $\tilde{\bm  T}$ defines on a whole $\mathbb{M}_4$
unit time measure. Decomposition of 4D objects onto its time-like and 3D
space-like projections can be carried out with the help of  {\it projection
operators}. They are constructed through the  all possible tensor powers of
$\tilde{\bm  T}$  and of  metrics of local space-like section ${\bm H}$,
defined by equation: $$ (\ ,\ )=\tilde{\bm  T}\otimes\tilde{\bm  T}-{\bm H}. $$

Particular interest has the case, when the form  $\tilde{\bm T}$ admits global
decomposition of $\mathbb{M}_4$ onto time and space. Necessary and sufficient
condition for the decomposition is {\it integrability of the form} $\tilde{\bm
T}$: $${\bm d}\tilde{\bm T}\wedge\tilde{\bm  T}=0, $$ where ${\bm d}$ ---
external differential. If it is valid, then differential equation $\tilde{\bm
T}=0$ has a solution $v(X)=t={\rm const}$, defining space-like hypersurface of
simultaneousity $\mathbb{V}_3^t$, to which coordinate time  $t$ corresponds.
This time is determined up to an arbitrary monotonic function $\varphi:\
t'=\varphi(t).$ At the surface  $\mathbb{V}_3^t$, which is, generally speaking,
Riemannian manifold with metrics $-{\bm H}$,  any 3D coordinates and their
points transformations $x'=x'(x)$ are admissible.

By analogy to bodies of classical mechanics, let us call continuous
non-negative real function $\varrho$: $\mathbb{M}_4\to\mathbb{R}$ {\it
4-density of a mass.} Then {\it 4-body} will be connected (in affine topology
of
$\mathbb{A}_4$) closed set $V\in\mathbb{M}_4$, for which $\varrho\neq0$. {\it
4-volume} of a 4-body  is the integral: $$ {\cal V}=\int\limits_{V}1,$$ and
{\it 4-mass} of the body
--- integral with the weight $\varrho$: $${\cal
M}=\int\limits_{V}\varrho.$$ The both integrations are carried out with
standard measure on $\mathbb{M}_4$: $$d{\rm vol_4}=\sqrt{-\Delta}dX^0\wedge
dX^1\wedge dX^2\wedge dX^3,$$ where $\Delta$ --- determinant of Minkowski
metrics. It will become  clear from the  following consideration, that in a
static world of 4-bars, 4-mass doesn't describe their inertial properties, but
sho\-uld be related to the elastic ones.

Having integrable  $\tilde{\bm  T}$-form and  4-body, one can consider  usual
{\it 3-bodies} as closed sets $V\cap\mathbb{V}_3^{t}=\cup v_i^t$, which give
non-vanishing restrictions: $\varrho|_{\mathbb{V}_3^t}\equiv\rho^t(x)>0$. It is
natural to call the restriction $\rho^t$ {\it 3-density of a mass}. So defined,
3-densities and 3-bodies correspond to ones  in classical Newton mechanics,
with the difference, that the first are depend on a choice  of $\tilde{\bm
T}$-form. To be short we'll not show this dependence explicitly. Using
decomposition of the  form of 4-volume: $d{\rm vol_4}=\tilde{\bm  T}\wedge {\rm
vol_3^t}$, one can calculate integral values of  {\it 3-volume} and  {\it
3-mass}:
\begin{equation}\label{mass}
\nu(t)=\int\limits_{v^t}1;\ \ m(t)=\int\limits_{v^t}\rho^t(x).
\end{equation}
From the 4D viewpoint, bodies, considered as different in  $\mathbb{V}_3^t$
(i.e. satisfying  $v_i^t\cap v_j^t=\varnothing$), can be different  sections of
the  same 4-body in $\mathbb{M}_4$ (if in the past or future they satisfy
condition $v_i^{t'}\cap v_j^{t'}\neq\varnothing$).

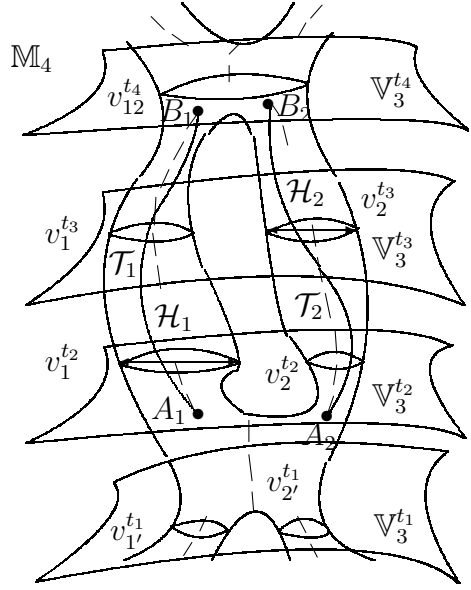
\begin{figure}[htb]
\centering \unitlength=1.00mm \special{em:linewidth 0.4pt}
\linethickness{0.4pt} \unitlength=1.3mm \special{em:linewidth 0.4pt}
\linethickness{0.4pt}
\begin{picture}(51.17,60.67)
\bezier{160}(22.00,60.67)(28.67,52.33)(33.33,60.67)
\bezier{140}(16.33,56.67)(23.00,49.33)(17.67,43.00)
\bezier{140}(37.33,57.67)(32.00,51.67)(37.67,43.00)
\bezier{260}(17.33,42.67)(9.33,31.67)(20.33,14.33)
\bezier{250}(38.00,42.33)(44.33,28.67)(37.00,14.00)
\bezier{70}(24.67,47.33)(27.67,51.33)(29.00,47.33)
\bezier{120}(24.00,33.67)(21.00,42.33)(24.33,47.33)
\bezier{100}(24.33,33.33)(29.00,23.33)(27.33,23.00)
\bezier{170}(29.33,47.00)(31.33,29.33)(32.67,26.67)
\bezier{130}(20.67,13.67)(24.33,7.00)(16.00,3.67)
\bezier{130}(36.67,13.67)(34.00,6.00)(41.67,3.67)
\bezier{170}(25.00,3.67)(30.00,13.67)(33.00,3.67)
\bezier{360}(7.17,1.50)(43.17,8.17)(50.17,1.33)
\bezier{130}(7.00,1.50)(17.33,8.67)(15.00,11.50)
\bezier{300}(15.00,11.67)(27.33,17.67)(49.00,14.83)
\bezier{150}(49.17,14.83)(44.00,11.17)(50.33,1.33)
\bezier{60}(21.00,7.17)(25.33,8.50)(26.67,6.67)
\bezier{60}(20.83,7.17)(23.33,5.17)(26.50,6.67)
\bezier{60}(31.83,6.67)(34.17,9.00)(36.83,7.00)
\bezier{50}(31.83,6.50)(36.00,5.50)(37.00,6.50)
\bezier{60}(27.17,23.00)(24.33,21.50)(28.17,18.83)
\bezier{200}(28.33,18.67)(40.50,16.83)(33.00,26.50)
\bezier{360}(6.17,15.67)(45.17,19.50)(50.83,16.17)
\bezier{130}(6.00,15.67)(16.00,20.33)(14.00,25.50)
\bezier{300}(14.00,25.50)(43.83,29.00)(49.33,25.67)
\bezier{130}(49.67,25.67)(44.33,24.50)(51.00,16.17)
\bezier{110}(15.67,23.67)(21.17,26.83)(27.83,24.00)
\bezier{100}(15.67,23.67)(21.67,22.00)(27.83,23.83)
\bezier{50}(34.67,24.00)(37.50,26.00)(40.50,24.00)
\bezier{50}(34.67,24.00)(38.17,22.50)(40.50,24.00)
\bezier{360}(6.17,29.83)(36.67,35.67)(50.67,29.83)
\bezier{130}(6.00,29.83)(15.33,35.83)(13.83,41.17)
\bezier{300}(13.83,41.33)(44.17,45.33)(49.67,42.50)
\bezier{140}(49.83,42.50)(44.50,39.00)(50.83,29.83)
\bezier{80}(30.67,37.17)(35.33,39.83)(39.83,37.67)
\bezier{80}(30.67,37.17)(35.83,35.17)(40.00,37.67)
\bezier{70}(14.50,37.17)(19.83,39.17)(22.83,37.33)
\bezier{70}(14.50,37.00)(20.17,35.50)(23.17,37.17)
\bezier{360}(5.83,47.17)(47.33,50.17)(51.17,47.67)
\bezier{130}(5.67,47.17)(15.67,51.00)(13.83,55.67)
\bezier{260}(13.83,55.83)(42.17,58.50)(46.00,56.17)
\bezier{100}(46.00,56.17)(44.50,53.17)(51.17,47.83)
\bezier{130}(19.83,51.33)(26.17,54.83)(34.83,52.50)
\bezier{120}(20.00,51.33)(28.33,50.00)(34.83,52.33)
\emline{22.17}{4.17}{1}{23.00}{5.50}{2} \emline{23.83}{6.67}{3}{24.50}{8.17}{4}
\emline{35.83}{4.00}{5}{35.17}{5.33}{6} \emline{34.67}{6.50}{7}{33.83}{8.17}{8}
\emline{23.33}{18.83}{9}{22.50}{20.67}{10}
\emline{22.00}{21.67}{11}{21.33}{23.83}{12}
\emline{19.83}{31.50}{13}{19.50}{33.83}{14}
\emline{19.33}{34.83}{15}{19.00}{37.17}{16}
\emline{19.17}{38.00}{17}{19.33}{40.83}{18}
\emline{19.83}{41.83}{19}{20.67}{43.67}{20}
\emline{21.17}{44.67}{21}{22.17}{46.50}{22}
\emline{22.50}{47.83}{23}{23.67}{49.50}{24}
\emline{26.83}{52.67}{25}{26.83}{54.33}{26}
\emline{20.17}{58.00}{27}{18.83}{59.33}{28}
\emline{29.33}{8.67}{29}{29.17}{10.67}{30}
\emline{29.17}{11.83}{31}{28.83}{14.50}{32}
\emline{28.83}{15.83}{33}{28.83}{18.00}{34}
\emline{36.83}{18.50}{35}{37.50}{20.50}{36}
\emline{37.50}{21.67}{37}{37.83}{23.67}{38}
\emline{37.83}{24.83}{39}{37.67}{27.17}{40}
\emline{37.17}{28.50}{41}{36.67}{31.00}{42}
\emline{36.33}{32.17}{43}{35.83}{34.67}{44}
\emline{35.67}{35.83}{45}{35.33}{38.00}{46}
\emline{32.83}{46.17}{47}{32.33}{48.17}{48}
\emline{32.00}{49.17}{49}{31.33}{50.17}{50}
\emline{26.83}{55.50}{51}{28.00}{56.67}{52}
\emline{33.83}{57.17}{53}{34.67}{58.67}{54}
\emline{35.17}{59.17}{55}{36.33}{60.50}{56}
\emline{22.67}{56.33}{57}{21.00}{57.50}{58}
\bezier{100}(23.50,18.67)(14.83,28.50)(19.17,39.17)
\bezier{100}(19.17,39.17)(23.33,45.33)(23.67,49.67)
\bezier{150}(36.83,18.50)(42.17,26.33)(36.00,33.50)
\bezier{150}(36.00,33.83)(30.17,41.00)(31.00,50.17)
\put(21.33,24.00){\vector(1,0){6.50}} \put(21.33,24.00){\vector(-1,0){5.83}}
\put(35.33,37.50){\vector(1,0){4.50}} \put(35.33,37.50){\vector(-1,0){4.67}}
\put(43.83,7.17){\makebox(0,0)[cc]{$\mathbb{V}_3^{t_{1}}$}}
\put(43.67,20.50){\makebox(0,0)[cc]{$\mathbb{V}_3^{t_{2}}$}}
\put(43.67,35.50){\makebox(0,0)[cc]{$\mathbb{V}_3^{t_{3}}$}}
\put(43.50,51.83){\makebox(0,0)[cc]{$\mathbb{V}_3^{t_{4}}$}}
\put(6.67,55.00){\makebox(0,0)[cc]{$\mathbb{M}_4$}}
\put(16.50,6.67){\makebox(0,0)[cc]{$v_{1'}^{t_1}$}}
\put(32.50,11.67){\makebox(0,0)[cc]{$v_{2'}^{t_1}$}}
\put(9.83,24.00){\makebox(0,0)[cc]{$v_1^{t_2}$}}
\put(32.33,22.83){\makebox(0,0)[cc]{$v_2^{t_2}$}}
\put(9.83,37.17){\makebox(0,0)[cc]{$v_1^{t_3}$}}
\put(42.17,40.67){\makebox(0,0)[cc]{$v_2^{t_3}$}}
\put(16.33,51.17){\makebox(0,0)[cc]{$v_{12}^{t_4}$}}
\put(21.17,28.50){\makebox(0,0)[cc]{$\mathcal{H}_1$}}
\put(34.67,41.33){\makebox(0,0)[cc]{$\mathcal{H}_2$}}
\put(23.67,18.67){\circle*{1.00}} \put(36.83,18.50){\circle*{1.00}}
\put(23.67,49.67){\circle*{1.00}} \put(30.83,50.33){\circle*{1.00}}
\put(20.67,19.00){\makebox(0,0)[cc]{$A_1$}}
\put(21.50,49.50){\makebox(0,0)[cc]{$B_1$}}
\put(36.00,16.33){\makebox(0,0)[cc]{$A_2$}}
\put(33.50,50.00){\makebox(0,0)[cc]{$B_2$}}
\put(16.00,34.00){\makebox(0,0)[cc]{$\mathcal{T}_1$}}
\put(35.00,29.67){\makebox(0,0)[cc]{$\mathcal{T}_2$}}
\end{picture}
\caption{\small Fragment of a 4-body.  Corresponding 3-bodies $v^t_i$ are shown
in some reference frame. Time-like and space-like sizes of 4-bars (see
following section) are equal to  $\mathcal{T}_1$, $\mathcal{T}_2$,
$\mathcal{H}_1$, $\mathcal{H}_2$ correspondingly. }\label{fb4}
\end{figure}

Mapping of the pair --- form of time and 4-body --- into set of 3-bodies:
$\{\tilde{\bm T},V\}\to \{v^t\in\mathbb{V}_3^t\}$ induces mapping,
$t\in\mathbb{R}\to v^t$, which we call {\it motion} in $\mathbb{V}_3^t$. It is
important to note, that this motion  is illusion from the $\mathbb{M}_4$
viewpoint:  it is sequence of intersection traces of $\mathbb{V}_3^t$ with
immobile  in $\mathbb{M}_4$ 4-body\footnote{Pav\u si\u c calls it in
\cite{pav2} $\Sigma$-motion (in our  notation $\Sigma=\mathbb{V}_3$)}. One can
say, that  the set of 4-bodies in  $\mathbb{M}_4$ is  ``frozen'' history.
Fragment of  4-body is shown in Fig.\ref{fb4}.

\section{THIN 4-BARS}\label{thbar}

Theoretically, one can consider 4-bodies, having restricted over  $t$ support
in some reference frame $\tilde{\bm T}$: $$\mathop{\rm
supp}\limits_{t\in\mathbb{R}}\varrho(t,x)=T\subset\mathbb{R}.$$ From the
viewpoint of $\mathbb{V}_3$ these 4-bodies may be looked at as suddenly
appearing  and disappearing 3-bodies. Such behaviour of classical (non-quantum)
objects is in drastic contradiction with observable continuous dynamics of any
3-bodies in $\mathbb{V}_3$. So, to  avoid paradoxes and violation of classical
conservation laws (for example, energy), one can put $$\mathop{\rm
supp}\limits_{t\in\mathbb{R}}\varrho(t,x)=\mathbb{R},$$ that implies eternity
of mechanical Universe. However, cosmological reasons demand and our everyday
experience admits weaker condition on  the support $\varrho$ in
$\mathbb{M}_4$. Lets call {\it space-like size} of a $4$-body $$
\mathcal{H}\equiv\max\limits_{t;\ A,B\in
v^t}\left|\,\int\limits_{\Gamma_{AB}^-\subset v^t} \sqrt{{\bm H}(d{\bm x},d{\bm
x})}\right|, $$ where $\Gamma_{AB}^-$ --- space-like geodesic, connecting some
two  points $A,B$ of 3-body  $v^t$. Roughly speaking, space-like size
$\mathcal{H}$ is maximal space-like size, chosen  from all space sections of
the 4-body (see Fig.\ref{fb4}). By analogy,  {\it time-like size } of the
4-body will be called  $$ \mathcal{T}\equiv\max\limits_{A,B\in
V}\int\limits_{\Gamma_{AB}^+}\tilde{\bm T},$$ where  $\Gamma_{AB}^+$ ---
time-like geodesic, connecting  some points  $A$ and  $B$ of the 4-body (see
Fig.\ref{fb4}). Now we can formulate condition, which is necessary for
agreement of properties of 4-bodies world with  observable 3-bodies ones: {\it
since observable classical 3-bodies easily show its bounds in 3-space, but
don't show bounds in time (i.e. don't appear out of nothing and disappear in to
nothing), then for every real 4-body we should put}
\begin{equation}\label{thin}
\mathcal{T}\gg \mathcal{H}.
\end{equation}
 Note, that inspite of dependency of  $\mathcal{H}$ and $\mathcal{T}$ on
reference frame, the non-equality  is relativistic invariant, by the well known
properties of Lorentz transformations. Such 4-bodies, in terms of elasticity
theory can be called {\it thin 4-bars}.

\section{4-D ELASTICITY THEORY}\label{4strain}

Let's call  {\it strain of  4-body}  the diffeomorphism
\begin{equation}\label{def}
V\to V_{{\bm \Xi}},
\end{equation}
given by the smooth \footnote{Components ${\bm \Xi}\in C^k,\ k\ge4.$} {\it
displacement vector field} ${\bm \Xi}={\bm \Xi}(X),\ \{X\}\in V$. As usually we
define {\it strain tensor} ${\bm D}$: $$dS^2_{{\bm \Xi}}=dS^2+2{\bm
D}(dX,dX),$$ where
\begin{equation}\label{strain}
{\bm D}=\frac{1}{2}(\overrightarrow{\partial}\otimes{\bm \Xi}+{\bm
\Xi}\otimes\overleftarrow{\partial})+\frac{1}{2}(\partial'\otimes\partial'')({\bm
\Xi'},{\bm \Xi''}),
\end{equation}
where $\partial$ --- tensor partial derivative operators, arrows and primes
denote their action on functions. In standard 3D linear elasticity theory
second term in (\ref{strain}) is commonly omitted, assuming  small value of
relative strains: $\max|\partial\xi|\ll1$, for which Hooks law is
valid\footnote{This is impossible, for example,  for weak bending of a thin
plates, when the first linear over derivatives of displacement vector term is
generally absent.}. This condition is invariant under transformation of
Galilean group acting in events space of classical mechanics by the compactness
of its  subgroup $\mathrm{SO(3)}$. The similar form of 4D bodies strains small
value conditions:
\begin{equation}\label{small}
\max|\partial{\bm \Xi}|\ll1,
\end{equation}
is, generally speaking, invalid, since the group $\mathrm{SO(1,3)}$ is ---
noncompact and pseudorotations of  reference frame can violate it. However,
since we only  try to formulate lagrangian formalism of classical
non-relativistic mechanics in terms of elasticity theory,  we can restrict our
attention to pseudorotations near group unity. Such transformations, which are
in fact translations in space of velocities, leave the condition
({\ref{small}}) invariant.

Decomposition of a  {\it 4-density of elastic free energy} in linear elasticity
theory contains only quadratic over components ${\bm D}$ terms:
\begin{equation}\label{gfren}
F=\zeta{\bm D}^2+\frac{\lambda}{2}\left[\mathop{\rm Tr}{\bm D}\right]^2,
\end{equation}
where $\zeta,\lambda$ --- phenomenological elastic {\it Lame coefficients}.
Stress 3-form $\ast{\bm \sigma}$, defining measure  $d\tilde{\bm F}$ of {\it
elastic 4-force 1-form}\footnote{We use notation $\ast{\bm \sigma}$ instead of
${\bm \sigma}$ to obtain standard relation (\ref{rule}). In coordinate form
$\ast{\bm \sigma}_{\alpha\beta\gamma\delta}={\bm E}_{\rho\beta\gamma\delta}{\bm
\sigma}^{\rho}_{\alpha}$, where $\bm E$ --- absolute antisymmetric Levi-Civita
tensor in $\mathbb{M}_4$.} $$\tilde{\bm F}(U)\equiv\int\limits_{\partial
U}\ast\tilde{\bm \sigma} =\int\limits_{U\subset V}{\bm d}\ast\tilde{\bm
\sigma}\equiv\int\limits_{U\subset V}d\tilde{\bm F},$$ where $\ast$ --- dual
conjugation in $\mathbb{M}_4$, $U$ --- arbitrary region inside 4-body $V$, can
be calculated by the formula \cite{land1}:
\begin{equation}\label{rule}
{\bm \sigma}=\frac{\partial F}{\partial {\bm D}}.
\end{equation}
 Boundary conditions at a
surface of the 4-body are:
\begin{equation}\label{bcond}
{\bm \sigma}|_{\partial V}=\tilde{\bm P},
\end{equation}
where $\tilde{\bm P}$ --- surface density of external 4-forces. Equation
(\ref{bcond}) implicitly assumes validity of  third Newton law for  the bodies
in $\mathbb{M}_4$, which, as it will be seen from the following consideration,
is the only mechanical law for 4-bars. This one, as it has been noted in
\cite{trus}, differing from the other two Newton laws, implies very general
assumptions about force function: it should only possess  an  additivity
property on the bodies of Universe: $\tilde{\bm F}(U_1\cup U_2)= \tilde{\bm
F}(U_1)+\tilde{\bm F}(U_2)$ for any $U_1,U_2$ satisfying $U_1\cap
U_2=\varnothing$.

\section{TWIST STRAINING OF A 4-BAR}\label{torsion}

\begin{figure}[htb]
\centering \unitlength=0.50mm \special{em:linewidth 0.4pt}
\linethickness{0.4pt}
\begin{picture}(270.00,70.00)(0.00,00.00)
\unitlength=0.50mm \special{em:linewidth 0.4pt} \linethickness{0.4pt}
\begin{picture}(137.00,87.00)
\bezier{260}(30.33,68.33)(13.33,65.00)(30.00,62.33)
\bezier{70}(30.33,62.33)(33.33,60.33)(28.00,60.00)
\bezier{140}(30.00,68.33)(40.33,69.67)(48.00,68.33)
\bezier{230}(48.00,68.33)(61.00,64.67)(45.67,63.33)
\bezier{90}(45.67,61.00)(40.00,62.33)(45.33,63.33)
\bezier{65}(27.00,60.00)(23.00,59.67)(27.00,58.67)
\bezier{72}(46.00,61.00)(50.33,61.00)(46.00,59.33)
\bezier{150}(27.67,58.33)(37.67,56.33)(45.67,59.00)
\emline{21.67}{65.33}{1}{21.67}{14.33}{2}
\emline{25.00}{59.33}{3}{25.00}{10.67}{4}
\emline{48.00}{60.67}{5}{48.00}{11.00}{6}
\emline{54.00}{65.67}{7}{54.00}{14.67}{8}
\bezier{40}(21.67,14.67)(22.67,12.33)(25.00,12.67)
\bezier{60}(54.00,15.00)(52.67,12.33)(48.00,12.33)
\emline{59.00}{38.00}{9}{78.33}{38.00}{10}
\emline{78.33}{38.00}{11}{72.67}{40.00}{12}
\emline{78.67}{38.00}{13}{72.33}{36.00}{14}
\bezier{260}(21.67,39.00)(36.67,43.00)(54.33,38.67)
\bezier{60}(21.67,39.00)(24.33,36.67)(27.33,37.00)
\bezier{70}(54.00,38.67)(51.00,36.33)(46.00,36.67)
\bezier{40}(25.00,35.33)(28.33,35.67)(27.33,36.67)
\bezier{40}(45.67,36.67)(45.00,35.33)(48.00,35.00)
\bezier{180}(25.00,35.00)(38.00,33.00)(48.00,35.00)
\bezier{260}(21.67,14.00)(38.67,19.33)(54.00,14.67)
\bezier{25}(25.00,12.33)(28.00,12.00)(28.33,12.00)
\bezier{25}(48.00,12.67)(47.00,11.67)(45.33,12.33)
\bezier{30}(25.00,12.67)(28.33,13.00)(28.33,12.00)
\bezier{40}(25.00,10.67)(29.00,11.33)(28.33,11.67)
\bezier{25}(47.67,11.00)(45.00,11.67)(45.33,12.33)
\emline{24.33}{63.67}{15}{32.67}{68.67}{16}
\emline{30.00}{62.33}{17}{40.00}{69.00}{18}
\emline{31.33}{57.67}{19}{45.67}{68.67}{20}
\emline{39.33}{57.67}{21}{43.33}{61.33}{22}
\emline{45.00}{63.33}{23}{50.67}{67.67}{24}
\emline{27.00}{37.00}{25}{33.33}{40.67}{26}
\emline{31.00}{34.33}{27}{40.00}{40.67}{28}
\emline{38.00}{34.00}{29}{45.67}{40.67}{30}
\emline{47.33}{37.00}{31}{50.00}{40.00}{32}
\emline{27.00}{13.00}{33}{30.33}{16.00}{34}
\emline{31.33}{9.33}{35}{38.67}{16.67}{36}
\emline{38.67}{8.67}{37}{44.67}{16.33}{38}
\emline{47.00}{12.33}{39}{49.33}{16.00}{40}
\bezier{180}(25.00,10.67)(36.00,6.67)(48.00,11.00)
\bezier{300}(86.67,15.33)(101.67,23.67)(119.67,15.33)
\bezier{30}(86.67,13.33)(85.00,14.33)(86.67,15.67)
\bezier{30}(120.33,15.00)(122.00,13.67)(120.33,13.33)
\bezier{80}(90.33,10.67)(93.00,13.67)(87.00,13.33)
\bezier{60}(120.00,13.00)(115.00,13.00)(117.00,10.33)
\bezier{80}(90.00,10.67)(85.33,10.67)(89.67,8.33)
\bezier{50}(117.33,9.67)(120.00,8.33)(117.00,7.67)
\bezier{200}(90.00,8.33)(104.33,5.33)(116.67,7.33)
\bezier{100}(91.33,41.00)(84.33,39.67)(89.33,37.33)
\bezier{170}(90.00,37.00)(102.00,31.67)(111.00,31.00)
\bezier{100}(112.00,31.00)(118.00,30.33)(113.67,35.33)
\bezier{50}(89.33,43.33)(93.33,42.00)(91.67,41.33)
\bezier{50}(113.67,36.00)(112.33,38.33)(116.67,37.00)
\bezier{100}(89.00,43.67)(86.67,47.67)(93.67,47.67)
\bezier{100}(116.67,37.33)(122.67,38.33)(119.33,42.33)
\bezier{200}(94.00,47.67)(107.67,47.67)(119.00,42.67)
\bezier{120}(108.33,56.33)(117.33,52.00)(118.33,58.00)
\bezier{80}(103.00,55.00)(103.00,59.33)(108.00,56.67)
\bezier{100}(118.33,58.33)(119.00,67.67)(116.67,71.33)
\bezier{200}(116.00,71.67)(108.33,79.33)(93.33,77.67)
\bezier{120}(93.00,77.67)(85.67,77.00)(89.00,69.67)
\bezier{60}(84.33,68.00)(89.00,66.33)(89.33,69.33)
\bezier{90}(83.67,68.33)(80.33,72.00)(79.67,65.67)
\bezier{80}(103.00,56.33)(103.33,51.67)(98.00,53.67)
\bezier{160}(79.67,65.33)(87.33,57.67)(97.33,54.00)
\bezier{230}(79.67,65.67)(86.67,56.00)(87.00,39.00)
\bezier{240}(87.33,9.67)(90.67,26.00)(87.00,39.00)
\bezier{130}(85.33,14.33)(88.67,28.33)(88.33,31.33)
\bezier{50}(88.00,74.33)(88.00,69.67)(87.00,67.33)
\bezier{140}(118.33,57.00)(116.33,49.67)(120.67,39.67)
\bezier{200}(121.00,14.00)(123.33,23.67)(120.33,40.33)
\bezier{200}(115.33,32.33)(114.33,21.67)(118.33,8.67)
\bezier{100}(115.67,33.67)(113.33,38.33)(108.67,44.33)
\bezier{100}(108.67,44.67)(104.67,49.33)(103.00,55.00)
\bezier{250}(81.00,64.00)(93.00,62.33)(105.00,77.33)
\bezier{260}(87.00,59.33)(101.33,59.67)(112.00,74.67)
\bezier{200}(95.67,54.67)(112.67,62.00)(118.00,69.00)
\bezier{70}(111.33,55.00)(116.00,57.00)(118.33,60.33)
\bezier{100}(88.67,71.00)(97.67,75.00)(97.33,78.00)
\bezier{100}(90.00,37.00)(95.67,39.67)(95.33,47.67)
\bezier{120}(96.33,34.33)(98.33,34.33)(102.33,47.00)
\bezier{130}(101.67,32.67)(109.33,39.33)(109.33,46.00)
\bezier{120}(108.00,31.33)(116.33,39.33)(116.00,43.67)
\bezier{90}(91.67,8.00)(94.67,13.67)(92.00,17.33)
\bezier{120}(98.00,7.00)(101.00,7.33)(99.33,19.33)
\bezier{100}(105.00,6.67)(108.00,12.67)(107.33,19.00)
\bezier{90}(113.33,7.00)(115.33,10.00)(115.00,17.33)
\bezier{180}(45.67,13.67)(28.33,16.33)(27.00,21.33)
\bezier{200}(26.67,21.67)(25.33,26.33)(45.67,26.33)
\bezier{400}(39.00,26.00)(63.00,28.67)(40.33,38.00)
\bezier{460}(40.00,38.00)(13.00,52.33)(40.00,52.33)
\bezier{500}(37.00,65.33)(66.67,55.33)(34.67,52.00)
\bezier{650}(45.67,19.33)(6.67,34.00)(46.33,31.33)
\bezier{340}(47.00,31.33)(62.33,32.67)(38.33,45.33)
\bezier{340}(38.00,45.33)(18.00,57.00)(36.67,59.67)
\emline{38.33}{22.33}{41}{42.00}{22.33}{42}
\emline{38.33}{22.33}{43}{40.33}{19.33}{44}
\emline{27.67}{20.33}{45}{31.33}{19.33}{46}
\emline{27.33}{20.33}{47}{29.00}{16.67}{48}
\emline{28.00}{46.00}{49}{31.67}{45.33}{50}
\emline{28.00}{46.00}{51}{30.00}{41.67}{52}
\emline{32.00}{49.33}{53}{35.33}{48.67}{54}
\emline{31.67}{49.33}{55}{33.67}{45.33}{56}
\emline{42.33}{53.00}{57}{38.67}{54.33}{58}
\emline{42.33}{53.00}{59}{39.00}{51.00}{60}
\emline{54.00}{53.00}{61}{85.33}{53.00}{62}
\emline{117.67}{53.00}{63}{137.00}{53.00}{64}
\emline{137.00}{53.00}{65}{113.67}{24.67}{66}
\emline{113.67}{24.67}{67}{0.00}{24.67}{68}
\emline{0.00}{24.67}{69}{21.67}{49.67}{70}
\put(12.67,30.67){\makebox(0,0)[cc]{$\mathbb{E}_3^t$}}
\put(70.67,44.67){\makebox(0,0)[cc]{$V\to V_{\bm\Xi}$}}
\put(44.00,47.67){\makebox(0,0)[cc]{$\bm\Xi$}}
\emline{37.08}{59.26}{71}{37.08}{54.23}{72}
\emline{37.08}{50.99}{73}{37.08}{46.77}{74}
\emline{37.08}{44.34}{75}{37.08}{40.29}{76}
\emline{37.08}{30.40}{77}{37.08}{26.35}{78}
\emline{37.08}{22.94}{79}{37.08}{19.70}{80}
\emline{37.05}{33.17}{81}{37.05}{37.90}{82}
\emline{37.05}{17.98}{83}{37.05}{13.69}{84}
\emline{37.05}{61.25}{85}{37.05}{66.12}{86}
\emline{101.69}{8.43}{87}{101.69}{14.06}{88}
\emline{101.69}{17.47}{89}{101.69}{23.95}{90}
\emline{101.69}{27.87}{91}{101.69}{34.18}{92}
\emline{101.69}{38.10}{93}{101.69}{43.39}{94}
\emline{101.69}{46.97}{95}{101.69}{52.94}{96}
\emline{101.69}{56.69}{97}{101.69}{62.48}{98}
\emline{101.52}{65.55}{99}{101.52}{70.50}{100}
\put(47.00,73.00){\makebox(0,0)[rc]{$\Phi(x)=0$}}
\emline{37.00}{13.67}{101}{30.50}{6.50}{102}
\emline{30.50}{6.50}{103}{31.50}{8.67}{104}
\emline{30.50}{6.67}{105}{32.83}{8.00}{106}
\emline{37.00}{13.67}{107}{50.83}{13.67}{108}
\emline{50.83}{13.67}{109}{49.17}{14.33}{110}
\emline{50.67}{13.67}{111}{49.00}{13.17}{112}
\emline{37.17}{37.33}{113}{30.83}{32.33}{114}
\emline{30.83}{32.33}{115}{31.83}{33.83}{116}
\emline{30.83}{32.33}{117}{33.33}{33.17}{118}
\emline{37.00}{37.17}{119}{50.83}{37.17}{120}
\emline{50.83}{37.17}{121}{48.50}{37.83}{122}
\emline{50.50}{37.17}{123}{48.67}{36.67}{124}
\emline{37.00}{64.33}{125}{29.00}{58.50}{126}
\emline{29.00}{58.50}{127}{30.17}{60.00}{128}
\emline{28.83}{58.33}{129}{31.00}{59.17}{130}
\emline{37.00}{64.17}{131}{50.33}{64.17}{132}
\emline{50.33}{64.17}{133}{48.00}{64.83}{134}
\emline{50.17}{64.17}{135}{47.83}{63.83}{136}
\emline{101.78}{11.34}{137}{99.65}{1.75}{138}
\emline{99.65}{1.75}{139}{99.65}{5.66}{140}
\emline{99.65}{1.93}{141}{101.42}{5.13}{142}
\emline{101.78}{11.34}{143}{115.62}{11.34}{144}
\emline{115.62}{11.34}{145}{111.89}{12.40}{146}
\emline{115.44}{11.34}{147}{111.72}{10.45}{148}
\emline{101.78}{40.63}{149}{95.21}{30.86}{150}
\emline{95.21}{30.86}{151}{96.10}{34.06}{152}
\emline{95.21}{31.04}{153}{97.69}{32.99}{154}
\emline{101.78}{40.80}{155}{112.96}{38.14}{156}
\emline{112.96}{38.14}{157}{111.01}{39.74}{158}
\emline{113.14}{37.96}{159}{110.30}{37.79}{160}
\emline{101.60}{65.48}{161}{78.88}{58.91}{162}
\emline{78.88}{58.91}{163}{81.90}{60.86}{164}
\emline{78.70}{58.91}{165}{83.14}{58.91}{166}
\emline{101.60}{65.48}{167}{108.88}{57.13}{168}
\emline{108.88}{57.13}{169}{107.63}{60.15}{170}
\emline{108.88}{57.13}{171}{106.04}{58.73}{172}
\emline{101.67}{65.00}{173}{108.33}{83.00}{174}
\emline{108.33}{83.00}{175}{105.33}{79.67}{176}
\emline{108.33}{83.00}{177}{108.00}{79.00}{178}
\emline{101.67}{40.67}{179}{105.00}{57.00}{180}
\emline{105.00}{57.00}{181}{105.33}{52.33}{182}
\emline{105.00}{56.33}{183}{103.33}{53.67}{184}
\emline{101.67}{11.33}{185}{103.33}{30.67}{186}
\emline{103.33}{30.67}{187}{102.00}{26.67}{188}
\emline{103.33}{30.67}{189}{103.67}{26.67}{190}
\end{picture}

\end{picture}
\caption{\small Twist strain of 4-bar. Coordinate system and form of time are
consistent with each other. In unstrained 4-bar state (from the left), 3-body
is in a rest. After straining (from the right) 3-body is  twisted and curved,
while inertia line remains rectilinear. }\label{f2}
\end{figure}
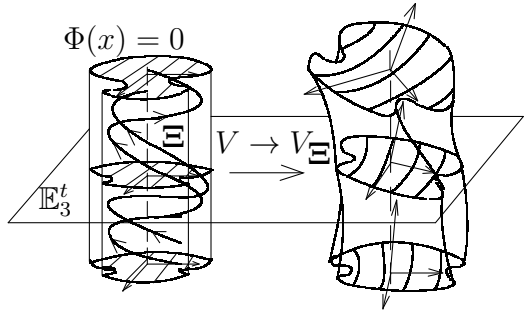

Let's consider a thin bar, occupying a  region $V\in\mathbb{M}_4$  in
unstrained case, given in Cartesian coordinates  $\{ X\}$ by the following way:
$$\mathop{\rm supp}\varrho=V= v\times[X_1^0;X_2^0],$$
---  where $v$ --- 3D set lying inside 2-surface, defined by some  equation
$\Phi(x)=0$.  Let's take form of time $\tilde {\bm T}=dX^0$, then
$\mathbb{V}_3^t=\mathbb{E}_3^t$, $v^t=V\cap\mathbb{E}_3^t=v=\mathop{\rm
const},$  and equation $\Phi(x)=0$ describes 3-space form  of some 3-body,
observed in reference frame $\tilde{\bm T}$ in a rest (see Fig.\ref{f2}). By a
parallel  transition in $\mathbb{E}_3$, one can put origin of space part of
coordinate system at the mass center ${\bm x}_0$ of the 3-body.

Let's call {\it twist strain } the special kind of diffeomorphism  (\ref{def}),
given  by the following displacement vector field: $${\bm \Xi}(X)={\bm
\Xi}(t,{\bm x})=(\Xi_0,\hat\mathrm{O}{\bm x}), $$ where $\hat\mathrm{O}$ ---
$t$-dependent 3D orthogonal matrix, defining twist angle of 3-body of the 4-bar
at arbitrary moment $t$. Its time derivative can be expressed through the {\it
twist 3-vector} ${\bm\omega}$ by the following expression: $$ \frac{d\hat
O}{dt}=[{\bm \omega},\ ], $$ where $[,]$ --- ordinary 3D vector product. Within
standard elasticity theory it has the meaning of twist angle of a bar per unit
of its length, while in 4-bars physics $\bm\omega(t)$ is, in fact, angular
velocity of 3-body, induced by the 4-bar twist strain \footnote{For convenience
we include light velocity $c$ into unit measure of time in all 4D expressions
and return it to apparent kind in all final results for comparing with
classical dynamics}. The component $\Xi_0=\Xi_0({\bm \omega},t,{\bm x}).$

Weak twist condition  $$\omega\mathcal{H}\ll1,$$ which we postulate now, and
which means in terms of SR slow (non-relativistic) rotation, let's expand
$\Xi_0$ in the row over powers of  ${\bm \omega}$: $$ \Xi_0={\bm \psi}_1({\bm
\omega})+{\bm \psi}_2({\bm \omega},{\bm \omega})+ {\bm \psi}_3({\bm
\omega},{\bm \omega},{\bm \omega})+\dots ,$$ where $${\bm
\psi}_n=\frac{\partial^n_{{\bm \omega}}\Xi_0}{n!}\left|_{{\bm
\omega}=0}\right.$$
--- {\it twist functionals} and lets then restrict our consideration  by the first term of the decomposition
(which we'll notate without index ``1''). Note, that twist functionals describe
space deformation of  3D sections (3-bodies) of the 4-bar and their
displacement along $t$-axe.

Non-zero components of linearized strain tensor (\ref{strain}) have the form:
\begin{equation}\label{nonz}
{\bm D}_{00}=(\dot{\bm \psi},{\bm \omega})+({\bm \psi},\dot{\bm \omega});\ \
{\bm D}_{0i}=\frac{1}{2}(({\bm \omega},\partial_i{\bm \psi})+[{\bm \omega},{\bm
x}]_i),
\end{equation}
where dot denotes time derivative, $\partial_i\equiv\partial/\partial x^i$.
Energy density   (\ref{gfren}) of twist strain then takes the form:
\begin{equation}\label{torfren}
 F_{\rm
tw}=(\zeta+\frac{\lambda}{2}){\bm D}_{00}^2-\zeta{\bm D}_{0i}^2=\frac{{\bm
J}_1({\bm \omega},{\bm \omega})}{2}+{\bm J}_2({\bm \omega},\dot{\bm \omega})+
\frac{{\bm J}_3(\dot{\bm \omega},\dot{\bm \omega})}{2},
\end{equation}
where tensors ${\bm J}_1,{\bm J}_2,{\bm J}_3$ are defined by the expressions:
$$ {\bm J}_1=2(\zeta+\frac{\lambda}{2})\dot{\bm \psi}\otimes\dot{\bm \psi}
-\frac{\zeta}{2} ((\partial{\bm \psi},\otimes\partial)_3{\bm \psi}+[{\bm x},
\overrightarrow{\partial}]\dot\otimes{\bm \psi}+{\bm \psi}\otimes[{\bm
x},\overleftarrow{\partial}] +{\bm x}^2(\ ,\ )_{3}-{\bm x}\otimes{\bm x}); $$
$$
 {\bm
J}_2=2(\zeta+\frac{\lambda}{2})\dot{\bm \psi}\otimes{\bm \psi};\ \ \   {\bm
J}_3=2(\zeta+\frac{\lambda}{2}){\bm \psi}\otimes{\bm \psi}. $$ Here $(\ ,\
)_3\equiv{\bm H}$ in our fixed reference frame. From the viewpoint of classical
mechanics the bilinear forms ${\bm J}_2({\bm \omega},\dot{\bm \omega}),\ {\bm
J}_3(\dot{\bm \omega},\dot{\bm \omega})$, and also the first term  in  ${\bm
J}_1({\bm \omega},{\bm \omega})$ describe relativistic corrections to the
kinetic rotational energy of a 3D solid; three addends in  ${\bm J}_1$,
including space derivatives of ${\bm \psi}$ describe ordinary elastic strain of
3-body, generated by rotation, and two remaining addendums in ${\bm J}_1$
determine kinetic energy density  of absolutely rigid 3-body. Really, for such
body ${\bm \psi}\equiv0$ and we get: $$ \mathfrak{F}_{\rm tw}|_{{\bm
\psi}\equiv0}=\int\limits_{V}F_{\rm tw }|_{{\bm \psi}\equiv0}\, dt\wedge d{\rm
vol}_3= -\int\limits_{V}\frac{\zeta}{2}{\bm J}_1({\bm \omega},{\bm
\omega})|_{{\bm \psi}\equiv0}\,dt\wedge d{\rm vol}_3 $$
\begin{equation}\label{equiv}
=\int\limits_{t_1}^{t_2}\left[\int\limits_{v}\frac{{\bm J}({\bm \omega},{\bm
\omega})}{2}\,d{\rm vol}_3\right]\wedge dt =\int\limits_{t_1}^{t_2}{\bm K}_{\rm
rot}\, dt=c\mathfrak{s}_{\rm rot},
\end{equation}
where   the equality $dt\wedge d{\rm vol}_3=-d{\rm vol}_3\wedge dt$  is taken
into account,  the identification
\begin{equation}\label{eq1}
\zeta=\rho c^2=\varrho c^2
\end{equation}
is made and  definition of  inertia tensor of rigid body:
\begin{equation}\label{inert}
2{\bm K}_{\rm rot}={\bm J}\equiv\int\limits_{v}\rho({\bm x}^2(\ ,\ )_{3}-{\bm
x}\otimes{\bm x})\, d{\rm vol}_3
\end{equation}
is used.

Influence of an external volume twist 4-force can be accounted by the 4-density
of potential energy $\Upsilon(X)$. Corresponding addendum to the free energy
will take the form:
\begin{equation}\label{pot4}
\mathfrak{F}_{\rm ext}=\int\limits_{V}\Upsilon\, dt\wedge d{\rm
vol}_3=-\int\limits_{t_1}^{t_2}U\, dt,
\end{equation}
where
\begin{equation}\label{pot3}
U\equiv\int\limits_v\Upsilon\, d{\rm vol}_3
\end{equation}
--- 3D potential energy.

\section{WEAK BENDING OF A BAR (STRING)}\label{bend}

Within standard elasticity theory in case of  a weak bending and
stretch-contracting the superposition principle takes place: elastic energies
for this kinds of strain are  calculated independently and then they are
combined in a full variational functional. When motion of solids in classical
mechanics is considered from the viewpoint of 4D elasticity, one should take
into account the both kinds simultaneously, in order to provide accordance with
experiment. Really, it is well known, that equations of bars bending have a
forth order
 \cite[p.110]{land1}, while equations of rigid motion of solids --- second one.
Note, that  stretch-contracting elastic energy, in case of a small displacement
depends on generalized coordinates derivatives quadratically and provides
second order equations of motion of 3D-bodies. However, it gives, in fact,
trivial 3D dynamics, since 4-bar under this pure stretch-contracting strains
remains non-curved, while general accelerated motion of a 3-body implies
curving of its world tube. These arguments lead us to the consideration of  a
{\it weak bending of tense bar}. Here and below we consider tension with its
sign. Let's introduce more rigorous notions and definitions, generalizing
standard bars bending theory to 4D case.

\subsection{Relativistic kinematics\\
 of a weakly bent bar}

\begin{figure}[htb]
\centering \unitlength=1.1mm \special{em:linewidth 0.4pt} \linethickness{0.4pt}
\begin{picture}(270.00,70.00)(0.00,00.00)
\unitlength=1.4mm \special{em:linewidth 0.4pt} \linethickness{0.4pt}
\begin{picture}(39.00,53.67)
\bezier{60}(9.83,6.00)(12.83,8.00)(15.83,6.00)
\bezier{60}(9.83,6.00)(12.83,4.00)(15.83,6.00)
\bezier{440}(9.67,6.00)(7.83,38.00)(29.17,47.33)
\bezier{360}(15.83,6.00)(14.67,35.00)(31.33,42.00)
\bezier{60}(29.33,47.33)(28.67,43.83)(31.50,42.00)
\bezier{60}(29.33,47.50)(32.33,45.83)(31.67,42.00)
\emline{9.50}{13.83}{1}{9.50}{18.67}{2} \emline{9.50}{20.83}{3}{9.50}{25.50}{4}
\emline{9.50}{27.67}{5}{9.50}{32.17}{6} \emline{9.50}{34.17}{7}{9.50}{38.17}{8}
\emline{9.50}{40.50}{9}{9.50}{44.50}{10}
\emline{9.50}{45.50}{11}{9.50}{48.67}{12}
\emline{15.83}{13.83}{13}{15.83}{18.50}{14}
\emline{15.83}{20.83}{15}{15.83}{25.67}{16}
\emline{15.83}{27.83}{17}{15.83}{32.17}{18}
\emline{15.83}{34.33}{19}{15.83}{38.33}{20}
\emline{15.83}{40.67}{21}{15.83}{44.50}{22}
\emline{15.83}{45.67}{23}{15.83}{48.50}{24}
\bezier{60}(9.50,48.67)(12.50,47.17)(15.83,48.67)
\bezier{60}(9.50,49.00)(12.33,50.50)(15.83,49.17)
\emline{12.67}{6.17}{25}{12.67}{53.67}{26}
\emline{12.67}{53.67}{27}{11.83}{50.67}{28}
\emline{12.67}{53.67}{29}{13.67}{50.50}{30}
\emline{12.67}{6.33}{31}{7.00}{1.67}{32}
\emline{7.00}{1.67}{33}{8.33}{4.00}{34} \emline{7.00}{1.50}{35}{9.50}{2.50}{36}
\emline{12.67}{6.33}{37}{33.33}{6.33}{38}
\emline{33.33}{6.33}{39}{30.33}{7.33}{40}
\emline{33.50}{6.33}{41}{30.33}{5.33}{42}
\bezier{36}(14.50,34.17)(14.67,29.50)(19.33,30.17)
\bezier{60}(14.50,34.00)(19.83,32.83)(19.50,30.17)
\emline{16.83}{32.17}{43}{23.83}{32.17}{44}
\emline{23.83}{32.17}{45}{22.33}{32.83}{46}
\emline{23.67}{32.17}{47}{22.50}{31.50}{48}
\emline{16.83}{32.17}{49}{20.83}{27.67}{50}
\emline{20.83}{27.67}{51}{20.33}{29.17}{52}
\emline{21.00}{27.67}{53}{19.17}{28.50}{54}
\emline{16.67}{32.17}{55}{20.67}{37.33}{56}
\emline{20.67}{37.33}{57}{19.00}{36.33}{58}
\emline{20.50}{37.00}{59}{20.17}{35.33}{60}
\bezier{72}(15.83,6.33)(7.17,14.00)(11.00,19.33)
\bezier{120}(11.00,19.50)(19.67,23.17)(17.17,29.50)
\bezier{110}(17.17,29.67)(16.33,40.50)(19.17,38.33)
\bezier{90}(19.50,38.17)(24.33,35.33)(27.17,41.00)
\bezier{80}(27.17,41.17)(27.33,48.33)(30.00,47.00)
\emline{17.00}{32.17}{61}{27.17}{26.83}{62}
\emline{27.17}{26.83}{63}{25.67}{28.67}{64}
\emline{27.00}{26.83}{65}{24.67}{27.17}{66}
\emline{2.00}{39.50}{67}{33.17}{39.50}{68}
\emline{33.17}{39.50}{69}{39.00}{46.00}{70}
\emline{39.00}{46.00}{71}{31.00}{46.00}{72}
\emline{26.17}{46.00}{73}{15.83}{46.00}{74}
\emline{9.50}{46.17}{75}{7.83}{46.17}{76}
\emline{7.83}{46.17}{77}{2.00}{39.50}{78}
\bezier{32}(9.50,41.33)(11.50,43.50)(15.83,41.67)
\bezier{60}(9.50,41.67)(13.00,40.50)(15.83,41.50)
\bezier{80}(20.00,41.33)(26.00,43.50)(29.00,41.00)
\bezier{70}(20.00,41.50)(24.00,40.17)(28.83,41.17)
\put(18.33,41.67){\vector(-1,0){5.67}} \put(18.33,41.67){\vector(1,0){6.83}}
\put(17.00,32.17){\circle*{1.05}} \put(12.67,6.33){\circle*{1.05}}
\emline{12.67}{13.17}{79}{13.33}{10.33}{80}
\emline{12.67}{13.17}{81}{12.00}{10.33}{82}
\put(10.83,2.17){\makebox(0,0)[cc]{$x$}}
\put(31.67,3.83){\makebox(0,0)[cc]{$y$}}
\put(14.17,14.50){\makebox(0,0)[cc]{$\bm U_{o}$}}
\put(22.17,38.00){\makebox(0,0)[cc]{$\bm U$}}
\put(19.67,26.17){\makebox(0,0)[cc]{$\bm e_x(S)$}}
\put(24.17,33.83){\makebox(0,0)[cc]{$\bm e_y(S)$}}
\put(26.17,25.00){\makebox(0,0)[cc]{$\bm A$}}
\emline{12.67}{12.83}{83}{13.17}{16.67}{84}
\emline{13.17}{17.83}{85}{13.83}{21.50}{86}
\emline{14.00}{23.33}{87}{14.83}{27.00}{88}
\emline{15.33}{28.50}{89}{16.67}{31.83}{90}
\emline{23.17}{40.00}{91}{25.67}{41.67}{92}
\emline{26.83}{42.17}{93}{29.00}{43.67}{94}
\emline{17.00}{32.17}{95}{13.50}{30.83}{96}
\emline{13.50}{30.83}{97}{14.67}{32.17}{98}
\emline{13.50}{30.83}{99}{15.50}{30.67}{100}
\put(14.00,29.00){\makebox(0,0)[cc]{$\bm \omega$}}
\bezier{90}(21.50,32.17)(22.67,28.83)(14.83,31.33)
\put(6.50,41.33){\makebox(0,0)[cc]{$\mathbb{E}_3^{t_2}$}}
\put(21.50,2.83){\makebox(0,0)[cc]{$\mathbb{E}_3^{t_1}$}}
\put(18.67,43.83){\makebox(0,0)[cc]{$\delta$}}
\put(14.67,51.83){\makebox(0,0)[cc]{$t$}}
\end{picture}
\end{picture}
\caption{\small Pure 4-bending of a bar. In unstrained state (shown by dash
line) 3-body is in rest. After straining 3-body is twisted and accelerated.
Twist vector $\bm\omega$, 4-acceleration  $\bm A$ and space-like triad $\bm e$
--- are coplanar. $\bm U$ --- tangent to inertia line 4-velocity vector, $\delta$ ---
space-like bending deflectio.}\label{fbend}
\end{figure}
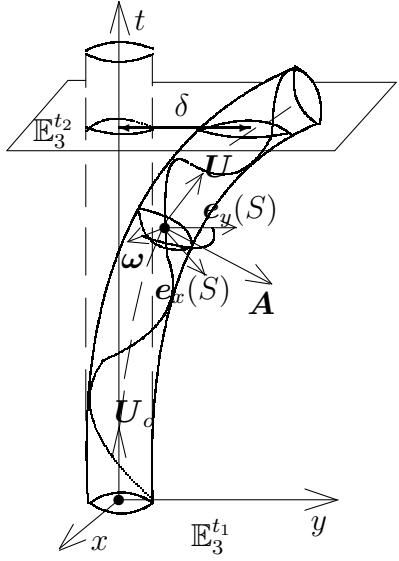

Let's pose a bar in its unstrained state in the same coordinate system, that
has been taken in previous section and lets consider it from the view point of
the same reference frame $\tilde{\bm  T}$ (see Fig.\ref{fbend}). General strain
of the thin bar can be described by orientation of orthonormal tetrad, rigidly
connected to points of the bar medium. We pose its origin at the inertia line
of the bar ${\bm X}_0(t)=\{t,{\bm x}_0(t)\}$) in every 3D section $t=const$
after strain.

In case of a pure bending (without global stretch-contracting) every 3-body,
defined by the reference frame $\tilde{\bm  T}$,  undergoes 4-rotation in
$\mathbb{M}_4$, which smoothly depends on length parameter $S$ along the bar.
Such rotation can be described by the Lorentzian  matrix
$\mathrm{L}(S)\in\mathrm{SO(1,3)}$, depending on six parameters: three of them
$\{\theta(S)\}$, describe pseudorotations (boosts), and three  ones
$\{\varphi(S)\}$ --- ordinary 3D rotations. Matrix  $\mathrm{L}$ satisfies the
following matrix relation:
\begin{equation}\label{inv}
\mathrm{L}^{\rm T }{\bm G}\mathrm{L}={\bm G},
\end{equation}
where  $\bm G$ --- pseudoeuclidian metrics, expressing conservation of 4-vector
norm under Lorentzian transformations. Form of inertia line of the strained
4-bar can be specified  by unit vector field ${\bm U}\equiv d{\bm
X}_0/dS=\dot{\bm X}_0$, which has physical sense of {\it 4-velocity of 3-body}.
If  ${\bm U}_0\equiv {\bm U}(0)$ is vector, taken at some (arbitrary but fixed)
initial location then for arbitrary $S$ we have: $${\bm U}(S)=\mathrm{L}(S){\bm
U}_0.$$ Lets consider vector field ${\bm U}$ at close point $S+dS$ of inertia
line: $${\bm U}(S+dS)\approx {\bm U}(S)+\dot{\bm U}dS=\mathrm{L}(S+dS){\bm U}_0
\approx(\mathrm{L}(S)+\dot \mathrm{L}dS){\bm U}_0,$$ and then
\begin{equation}\label{spin}
\dot{\bm U}\equiv{\bm A}=\dot\mathrm{L}(S){\bm
U}_0=\dot\mathrm{L}\mathrm{L}^{-1}{\bm U}(S),
\end{equation}
where  {\it acceleration 4-vector}  ${\bm A}$ is introduced. Linear operator
$\dot\mathrm{L}\mathrm{L}^{-1}\equiv{\bm \Omega}$ can be  called  {\it torsion
4-tensor} of the bar. It has a geometrical sense of linear connection:
(\ref{spin}) can be rewritten in the form $D{\bm U}=0$, where covariant
derivative along inertia line operator  is defined by relation: $D\equiv d-{\bm
\Omega}(\cdot\ ,\ )dS.$

4-torsion ${\bm \Omega}$, as it follows directly from its definition
(\ref{spin}), is linear form over derivatives $\{\dot\theta,\dot\varphi\}$.
Since  $\{\dot\theta\}\sim{\bm A}$, space-like vector ${\bm A}$ is responsible
for bending of the 4D-bar. More exactly, value $|{\bm A}|^2=-K^2$, where $K$
--- first curvature of the inertia line.
Then, weakness of a bending can be formulated through the following Lorentz
invariant integral relation:
\begin{equation}\label{little}
\int\limits_{S_1}^{S_2}K(S)dS\ll1\ \ \forall\ S_1<S_2.
\end{equation}
As it is well known from differential geometry \cite{eiz}, in neighborhood of
any point of smooth regular curve the difference between the curve and tangent
plane, defined by the pair  ${\bm U}$ and ${\bm A}$ is value $o(S^2)$, where
$S$ --- displacement of the point along curve. Consequently all high curvatures
can be neglected in calculation of local strain in linear elasticity theory.

The derivatives $\{\dot\varphi\}$  are proportional to the angular velocity
vector ${\bm \omega}$, whose relativistic representation has the form:
\begin{equation}\label{rotvel}
{\bm \omega}\equiv\ast({\bm \Omega}\wedge{\bm U})=\frac{1}{3!}{\bm E}(\ ,
{\bm\Omega},{\bm U})=\frac{1}{3!}{\bm\epsilon}(\ ,{\bm\Omega})\equiv\star{\bm
\Omega},
\end{equation}
where ${\bm \epsilon}\equiv {\bm E}(\ ,\ ,\ ,{\bm U})= \ast{\bm U}$
--- {\it Levi-Civita 3-tensor} in $\mathbb{V}_3$, "$\star$" --- dual conjugation in $\mathbb{V}_3$.
Let us remind that in case of ordinary bar when it is generally bent, the
bending strain is inevitably followed by twist one, considered  in previous
section.

Easy to see that the torsion 4-tensor  ${\bm \Omega}$ is  relativistic 4-bar
strain measure of a 4-bar and its components and first derivatives (see
(\ref{nonz})) are linearly connected
 with
strain tensor ${\bm D}$. Then  quadratic over ${\bm D}$ expression for a free
energy  (\ref{gfren}) can be written then in terms of quadratic combinations
of   ${\bm \Omega}$ and $\dot{\bm \Omega }$:
\begin{equation}\label{fren4}
F=\frac{1}{2}\mathcal{J}^{(I)}({\bm \Omega},{\bm \Omega})+
\frac{1}{2}\mathcal{J}^{(II)}(\dot{\bm \Omega},\dot{\bm \Omega})+
\mathcal{J}^{(III)}(\dot{\bm \Omega},{\bm \Omega}),
\end{equation}
where it is natural to call functionals  $\mathcal{J}^{I,II,III}$  {\it inertia
4-tensors}. For our purposes,  only the first term should be held, since every
time derivative implicitly  contains   multiplier $1/c$, and, consequently,
 the second and third terms can be related to relativistic corrections.
Let ${\bm \Omega}={\bm \Omega}_1({\bm A})+{\bm \Omega}_2({\bm \omega})$, where
${\bm \Omega}_{1,2}$ --- are corresponding linear functionals. After
substitution ${\bm \Omega}$ into  (\ref{fren4}), we get  $F$  as the following
quadratic form:
\begin{equation}\label{fren3}
F=\frac{1}{2}\mathcal{J}_1^{I}({\bm A}, {\bm A})+\mathcal{J}_2^{I}({\bm A},{\bm
\omega})+\frac{1}{2}\mathcal{J}_3^{I}({\bm \omega},{\bm \omega}).
\end{equation}
Third term, when  strain of 3-body is omitted, must be identical with
expression (\ref{torfren}) for a pure twist under ${\bm \psi}=0$, so
$\mathcal{J}_3^{I}$ is  3-density of inertia tensor ${\bm J}$ in (\ref{inert}).
Second term should be put zero by a time symmetry of elastic energy. In
classical language $\mathcal{J}_2^{I}\equiv0$ by $T$-invariance of its laws.

\subsection{Apparent kind of the functional $\mathcal{J}_1^{I}$}

For making calculations short, we'll generalize known 3D result for
$\mathcal{J}_1^I$ functional \cite[p.93]{land1}. Let ordinary bar under
external influence has curvature in plane $(z,x)$, where $z$ coincides with axe
of the bar in undeformed state. Then 3-density of free energy has the form:
\begin{equation}\label{fren1}
F=\frac{Ek^2x^2}{2},
\end{equation}
where  $E$ --- Young modulus, relating with Lame coefficients in 3D case by the
expression:
\begin{equation}\label{Young}
E=3\zeta-\frac{\zeta^2}{\lambda+\zeta},
\end{equation}
$k$ --- curvature of a middle line of the bar at the point. Note, that
$k^2=|{\bm k}|^2$, where  ${\bm k}=d{\bm \tau}/dl$ --- curvature vector, ${\bm
\tau}$
--- unit tangent to inertia line of the bar vector, $dl$ --- line element of the bar.
In case of general position of the plane of curving, expression (\ref{fren1})
can be generalized  by the following way:
\begin{equation}\label{gfren1}
F=\frac{E}{2}({\bm x}\otimes{\bm x})({\bm k},{\bm k}),
\end{equation}
Elastic energy per length unit is $$\frac{d\mathfrak{F}}{dl}=\frac{E}{2}{\bm
i}({\bm k},{\bm k}),$$ where $${\bm i}\equiv\int_s({\bm x}\otimes{\bm x})$$ ---
{\it inertia moment of the section of a bar} and integration  is carried out
over the cross section of the bar. Full bending energy is
$$\mathfrak{F}=\int\limits_{0}^{l}\frac{d\mathfrak{F}}{dl}dl.$$

For generalization to the case of 4-bars we take into account, that for
isotropic elastic body in  $\mathbb{M}_4$ (\cite{kok1,kok2}):
\begin{equation}\label{young}
E=\frac{8}{3}\zeta-\frac{4}{9}\frac{\zeta^2}{(\lambda+2\zeta/3)},
\end{equation}
then note, that role of ${\bm k}$ plays  4-acceleration  vector ${\bm A}$.
Finally we generalize expression for ${\bm i}$: $${\bm
I}=\int\limits_{v^t}({\bm X}\otimes{\bm X})\,d{\rm vol}^t_3.$$ So, we get
$$\mathcal{J}_1^I=\frac{E}{2}{\bm I}$$ and finally
\begin{equation}\label{frenb}
\mathfrak{F}_{b}=\int\limits_{S_1}^{S_2}\mathcal{J}_1({\bm A},{\bm A})\,
dt\wedge d{\rm vol}^t_3.
\end{equation}

In the following part of the paper expression  (\ref{frenb}) will be used only
for physical and geometrical restrictions analysis. Really, the part
(\ref{frenb}) of a full free energy, include second derivatives of displacement
vector and so, as it has been mentioned, equilibrium equations will have  forth
order. In the next subsection we'll construct "right" variational functional
$\mathfrak{F}_{\rm s}$, which quadratically depends on the first derivatives of
displacement vector components and includes  tension. For our purposes the
following (now obvious) fact is important: {\it since whole classical mechanics
should be  contained in $\mathfrak{F}_s$, we have the non-equality}
\begin{equation}\label{cond}
\mathfrak{F}_b\ll\mathfrak{F}_s,
\end{equation}
which, in fact,  redirects our consideration from a general 4-bars physics to
the more particular case of 4-strings.

\subsection{Bend energy of 4-string}

Let's  consider 4-string in equilibrium under influence of the forces ${\bm
T}_1=-{\bm T}_2$, applied at the  ends $v^{t_1}$ and $v^{t_2}$. This kind of
strain is called {\it simple stretching.}

\begin{figure}[htb]
\centering \unitlength=1.00mm \special{em:linewidth 0.4pt}
\linethickness{0.4pt}
\begin{picture}(270.00,70.00)(0.00,00.00)
\unitlength=0.8mm \special{em:linewidth 0.4pt} \linethickness{0.4pt}
\begin{picture}(53.33,72.33)
\bezier{40}(11.33,9.00)(12.67,7.00)(15.33,7.67)
\bezier{60}(11.00,9.00)(15.00,10.33)(15.67,8.00)
\bezier{250}(11.00,9.00)(21.67,17.67)(26.33,35.33)
\bezier{260}(15.67,8.00)(28.00,20.00)(30.33,35.67)
\bezier{300}(26.33,35.67)(33.00,62.33)(42.00,64.00)
\bezier{240}(30.33,36.33)(38.33,60.33)(43.33,59.67)
\bezier{40}(42.33,64.00)(41.33,61.33)(43.33,59.67)
\bezier{40}(42.33,64.00)(44.00,63.00)(43.33,60.00)
\emline{25.00}{59.67}{1}{25.00}{55.33}{2}
\emline{25.00}{52.67}{3}{25.00}{48.33}{4}
\emline{25.00}{45.00}{5}{25.00}{41.00}{6}
\emline{25.00}{38.00}{7}{25.00}{33.33}{8}
\emline{25.00}{30.67}{9}{25.00}{26.67}{10}
\emline{25.00}{24.00}{11}{25.00}{19.67}{12}
\emline{25.00}{17.00}{13}{25.00}{13.00}{14}
\emline{29.00}{12.67}{15}{29.00}{17.33}{16}
\emline{29.00}{20.00}{17}{29.00}{24.00}{18}
\emline{29.00}{27.00}{19}{29.00}{31.00}{20}
\emline{29.00}{33.67}{21}{29.00}{38.00}{22}
\emline{29.00}{41.00}{23}{29.00}{45.00}{24}
\emline{29.00}{48.00}{25}{29.00}{53.00}{26}
\emline{29.00}{55.00}{27}{29.00}{60.00}{28}
\bezier{40}(25.00,60.00)(27.00,61.33)(29.00,60.00)
\bezier{40}(25.00,60.00)(27.67,58.67)(29.00,60.00)
\bezier{40}(25.00,12.33)(27.33,13.67)(29.00,12.33)
\bezier{40}(25.00,12.33)(27.33,11.00)(29.00,12.33)
\emline{46.33}{55.67}{29}{38.33}{55.67}{30}
\emline{38.33}{55.67}{31}{43.00}{56.67}{32}
\emline{38.67}{55.67}{33}{43.00}{54.67}{34}
\emline{36.00}{51.67}{35}{43.00}{51.67}{36}
\emline{36.33}{51.67}{37}{40.33}{53.00}{38}
\emline{36.33}{52.00}{39}{40.00}{50.33}{40}
\emline{34.33}{47.33}{41}{41.67}{47.33}{42}
\emline{34.33}{47.33}{43}{38.33}{48.67}{44}
\emline{34.00}{47.33}{45}{38.33}{46.00}{46}
\emline{33.00}{44.00}{47}{43.00}{44.00}{48}
\emline{33.00}{44.00}{49}{37.33}{45.33}{50}
\emline{33.00}{44.00}{51}{37.33}{42.67}{52}
\emline{7.33}{14.00}{53}{15.67}{14.00}{54}
\emline{13.67}{15.33}{55}{16.33}{14.00}{56}
\emline{16.33}{14.00}{57}{13.00}{12.67}{58}
\emline{9.33}{18.67}{59}{19.33}{18.67}{60}
\emline{19.33}{18.67}{61}{16.33}{20.33}{62}
\emline{19.33}{18.67}{63}{16.00}{17.33}{64}
\emline{21.33}{22.67}{65}{13.00}{22.67}{66}
\emline{21.33}{22.67}{67}{17.33}{24.33}{68}
\emline{21.67}{22.67}{69}{16.33}{21.33}{70}
\emline{16.00}{27.33}{71}{23.67}{27.33}{72}
\emline{23.67}{27.33}{73}{19.67}{29.00}{74}
\emline{20.00}{26.00}{75}{23.67}{27.33}{76}
\emline{27.00}{60.00}{77}{27.00}{72.33}{78}
\emline{27.00}{72.33}{79}{25.67}{67.33}{80}
\emline{26.67}{72.33}{81}{28.67}{67.33}{82}
\emline{27.00}{12.33}{83}{27.00}{1.67}{84}
\emline{27.00}{1.67}{85}{25.67}{6.33}{86}
\emline{27.00}{2.00}{87}{28.67}{6.33}{88}
\emline{13.67}{8.33}{89}{7.67}{2.33}{90}
\emline{7.67}{2.33}{91}{9.00}{6.00}{92}
\emline{7.67}{2.67}{93}{11.67}{4.00}{94}
\emline{42.67}{62.33}{95}{52.33}{67.00}{96}
\emline{52.33}{67.00}{97}{48.00}{66.67}{98}
\emline{52.33}{67.00}{99}{50.00}{64.00}{100}
\put(46.67,49.33){\makebox(0,0)[cc]{$\bm F(X)$}}
\put(6.33,21.33){\makebox(0,0)[cc]{$\bm F(X)$}}
\put(31.33,69.67){\makebox(0,0)[cc]{$\bm T_1$}}
\put(31.33,4.67){\makebox(0,0)[cc]{$\bm T_2$}}
\put(4.67,4.00){\makebox(0,0)[cc]{$\bm T'_2$}}
\put(53.33,63.67){\makebox(0,0)[cc]{$\bm T'_1$}}
\end{picture}
\end{picture}
\caption{\small Bending of a tense string. Arrows show bending and stretching
forces. Bent string get extrastretching, which energy much more then purely
bending  one. }\label{fstr}
\end{figure}
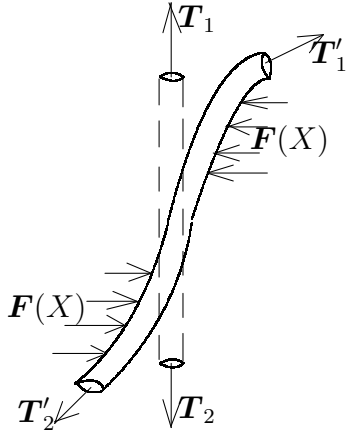

In ordinary bar deformation theory the following {\it Saint-Venant} principle
is commonly used: \cite[p.33]{rab}: {\it external forces distribution at the
ends of a bar directly influences on a stress picture only near the ends at the
distance about thickness of a bar.} In other words, far from ends, stress
tensor depends only on  integral values of forces (or its momenta), applied at
the ends, but doesn't depend on their local distributions there. Now we
postulate validity of the  principle in case of 4-strings, defering its
physical sense discussion to the Conclusion.

In the simple stretch case the only non-zero component of stress tensor
$\bm{\sigma}$ is
\begin{equation}\label{stretch}
{\bm \sigma}_{tt}(t)=T/\nu_t\equiv\Pi=E{\bm D}_{tt},
\end{equation}
where ${\bm D}_{tt}$ --- relative time-like stretching of the string.
Expression  (\ref{stretch}) can be viewed as definition of a Young modulus $E$.
Similarly, the expression for cross strain tensor components can be considered
as definition of {\it Poisson coefficient} $\sigma$:
\begin{equation}\label{poiss}
{\bm D}_{xx}={\bm D}_{yy}={\bm D}_{zz}=-\sigma{\bm D}_{tt},
\end{equation}
where for bodies in ${\mathbb{M}_4}$  $\sigma=\lambda/(3\lambda+2\zeta)$
(\cite{kok1,kok2}). Now if the stretched string is affected to bending, then,
generally speaking, its integral length will also vary: $S_0=|t_2-t_1|\ \to\
S$. Change of the string elastic tension energy under the bending is:
\begin{equation}\label{frens}
F_{s}=\frac{{\bm \sigma}_{tt}{\bm D}_{tt}}{2}=\frac{\Pi}{2}\frac{d\Delta
S}{dt},
\end{equation}
For the $S-S_0$, taking into account weakness of the bend\footnote{Rigorously
speaking, we use the  {\it smallness of an inclination} of the 4-string
relative to time coordinate  line in some reference frame $\tilde{\bm T}$, that
can be expressed by the non-equalities ${\bm v}^2\ll1$ at whole length of the
4-string. This condition is, obviously, sufficient, but not necessary for weak
bend (\ref{little}). In ordinary language of velocities and accelerations, we
consider nonrelativistic approximation, which doesn't follow from a
acceleration smallness of the 3-body. } we have: $$ \Delta
S=\int\limits_{S_1}^{S_2}\sqrt{({\bm U},{\bm U})}dS- S_0
=
\int\limits_{t_1}^{t_2}\sqrt{1-{\bm v}^2}\, dt-S_0\approx
-\frac{1}{2}\int\limits_{t_1}^{t_2}{\bm v}^2\, dt$$ and finally, for stretch
elastic energy density $$ F_{\rm s}=-\frac{\Pi}{4}{\bm v}^2. $$ Integrating it
over 4-volume of the string we get:
\begin{equation}\label{class}
\mathfrak{F}_{\rm s}=c\int\limits_{t_1}^{t_2}\frac{m{\bm v}^2}{2}\,
dt=c\mathfrak{s}_{\rm pr},
\end{equation}
--- classical progressive part of action for a solid, when
\begin{equation}\label{eq2}
\frac{\Pi}{2}=\rho c^2\ \  {\rm and}\ \ mc^2=\frac{1}{2}\int\Pi\, d{\rm
vol}^t_3.
\end{equation}
Comparing it with  (\ref{eq1}), we find
\begin{equation}\label{eq3}
\zeta=\frac{\Pi}{2}
\end{equation}
--- another "tuning relation" for elastic properties of the 4-string.
Potential progressive term can be introduced similarly to the section
\ref{torsion} (formulaes (\ref{pot4})-(\ref{pot3}).

At the end of the section lets clear those conditions, which provide identity
of the bar deformation  theory  and ordinary classical mechanics. For this
purpose we estimate energies $F_{b}$ and $F_{s}$. For the former, using
(\ref{frenb}), we have  $$ F_{b}\sim E\mathcal{H}^2|{\bm
A}|^2\sim\frac{E\mathcal{H}^2\delta^2}{\mathcal{T}^4}, $$ for the latter from
the (\ref{class}): $$
F_{s}\sim\Pi\left(\frac{\delta}{\mathcal{T}}\right)^2\sim\frac{T\delta^2}{\mathcal{H}^3\mathcal{T}^2},
$$ where $\delta={\rm max}_t\int_{t_1}^{t}\sqrt{-{\bm H}(d{\bm X}_{0},d{\bm
X}_0)}$
--- {\it bending deflectio} of  4-string, equal to maximal space-like displacement of  3-body
during the time from $t_1$ to $t_2$ (see Fig.\ref{fbend}).

Remembering condition (\ref{cond}), we get
\begin{equation}\label{est}
\frac{F_s}{F_b}\sim\frac{T}{E\mathcal{H}^3}\cdot\left(\frac{\mathcal{T}}{\mathcal{H}}\right)^2\gg1.
\end{equation}
Second multiplier in  (\ref{est}) is much more than unit by definition of a
thin bar (\ref{thin}) in \S\ref{thbar}. It means, that the first one
$T/E\mathcal{H}^3$ can not be much less then unit.  By order of magnitude the
denominator is the 4-force $T_{\rm cr}$, which should be applied to the string
to change its length by value about original one, assuming that Hooks law
remains valid. So the possibility of pure bending energy omitting in comparison
with stretching one for 4-strings (and consequently, validity of classical
Newton mechanics), lies in the fact, that   {\it tension of a 4-string is not
substantially different from  $T_{\rm cr}$.}

\section{CONCLUSION}\label{concl}

Let's summarize and discuss obtained  results and another physical and
mathematical analogies of classical mechanics with multidimensional
generalization of elasticity and  material strength theories.

\begin{enumerate}
\item
In standard approach within  SR we deal with world of events $\mathbb{M}_4$ and
world tubes of bodies, commonly used  as convenient way for visualizing of
their motion. In the proposed approach the world tube of 3-body is treated as
{\it unified physical object,} existing as  4-body (thin string) and possessing
4D elastic properties\footnote{This is impossible in events space of
 classical mechanics $\mathbb{R}\times\mathbb{E}^3$, since here we have two independent
 and incompatible metrics --- euclidian one and absolute form of time. So world tubes of 3-bodies
 in $\mathbb{R}\times\mathbb{E}^3$ are not physical, but purely formal objects.}.
Observer should understand it as {\it absolute (or objective) history of a some
3-body}. This 3-body is defined by the observers reference frame and its
absolute history transforms into {\it relative (or  subjective)} one, following
to observer's perception  (devices) motion from one simultaneousity
hypersurface to another.
\item
Previous investigation allow us to write the following relation between
elasticity  theory of strings and classical mechanics of solids:
\begin{equation}\label{analogy}
\mathfrak{F}_{\rm str}=\mathfrak{F}_{\rm str\ s}+\mathfrak{F}_{\rm str\ tw}=
c\mathfrak{s}_{\rm pr}+c\mathfrak{s}_{\rm rot},
\end{equation}
where $\mathfrak{F}_{\rm str\ s},\ \mathfrak{F}_{\rm str\ tw}$ --- elastic
4-energies of string extrastretching (induced by its bending) and twist and
$\mathfrak{s}_{\rm pr},\ \mathfrak{s}_{\rm rot}$ --- actions for solids
progressive  and rotational motions correspondingly. This expression is in
straightforward analogy to (\ref{act}) and (\ref{frengr}). We list the
assumptions to be accepted to obtain (\ref{analogy}) in the Table 1. The
1,2,5-th  lines are quite obvious. 4-th line restricts general theory of bars
deformation to strings one, where any 3D  section of a string is managed by
ordinary Newton equations. Acceleration de\-pen\-den\-ce of the classical
action can appear as a small correction, which can be connected to with modern
investigation on modified Newton dynamics (MOND) \cite{mond}. The lines 3-rd
and 6-th of the table can be treated as some kind of "tuning": elastic
properties of a string with respect to twist and stretch\-ing-con\-tract\-ing
are consistent, so that only one observable parameter --- mass 3-density $\rho$
appears in equations. From a general point of view, twist stiffness (rotational
inertia) and bending one (progressive inertia) could be independent and quite
different. Combining 3-th and 6-th lines and using (\ref{young}) and
(\ref{est}) after some transformations we get the following relation between 4D
elastic modulus: $\lambda/\zeta\sim1$, that is true for ordinary elastic
3-bodies.
\item
Important and interesting circumstance, revealed in the approach is that the
only postulates about forces and third Newton law validity are necessary, which
allow us to write 4D static equilibrium equations. Mechanical motion and all
involved notions have secondary nature: velocity and acceleration have purely
differentially-geometric origin and  mass has purely force-like one. Really,
classical action include the only integral parameter, which is "response from
the ends" ---  mass density of the 3-body. But according to (\ref{eq2})
\footnote{4-force has physical dimension of 3-energy.}, $mc^2={\rm T/2}$.
Classical  law of a mass conservation ($\dot m=0$ in (\ref{mass})) is then the
straightforward consequence of its force nature, since $m(t_1)=m(t_2)$ means
$T(t_1)=T(t_2)$ --- equilibrium condition of a string along $t$-direction.
\item
Forces, distributed on  time-like surface of a string are ordinary contact
forces, which we deal with in classical mechanics. By analogy with ordinary
3-strings, which satisfy Saint-Venant principle, its deformation  is
independent on forces distributions  far from the points of its application.
The deformation only depends on  integral force distribution characteristics:
total force ${\bm F}=\int d{\bm F}$  defines final bending angle of the
3-string and total force momentum ${\bm M}=\int{\bm x}\wedge d{\bm F}$
--- defines value of a constant twist of the 3-string (see.\cite[p.91,p.103]{land1}).
This fact, generalized to 4-strings physics becomes Galilean inertia principle:
{\it strongly tensed 4-string, at regions where there is no bending or twisting
forces, remains rectilinear and possesses constant twist}.
\item
It is easy to note, that second Newton law, say, for progressive motion
expresses static balance of external  bending forces and string elastic
reaction to the bending equal to $-m \ddot{\bm x}$ in terms of the approach.
Consequently, {\it inertia forces} acquire the sense of elastic reaction forces
to bending  (progressive inertia) and to twist (rotational inertia.)

Let's show, that second Newton law can be treated as the special (1D) case of
{\it Laplace formulae}:
\begin{equation}\label{laplace}
\Delta p=q\bar k\alpha,
\end{equation}
which connects under equilibrium normal pressure difference between each side
of elastic  2D-membrane with its local surface tension $\alpha$ and local
curvature arithmetical mean $\bar k=(k_1+k_2)/2$, where $k_i=1/R_i$ --- main
curvatures of the membrane. Geometrical dimensionless factor $q=2$ for the case
of 2D membrane in 3D space. If we change in (\ref{laplace}) bending pressure
$p\to {\bm f}$ by the following correspondence $$d{\bm f}_{\rm b}=p\, d{\bm s}\
\longrightarrow\ d{\bm F}_{\rm b}= {\bm f}dt,$$ where ${\bm f}_{\rm b},{\bm
F}_{\rm b}$ --- 3D and 4D bending forces, $\bm f$ --- 3D ordinary one, $d{\bm
s}$ --- normal element of 2D membrane, then $\alpha\to T$ by the correspondence
$$ du=\alpha ds\ \longrightarrow\  dU=Tdt=2mc^2dt, $$ where $du,dU$ --- 2D and
4D  elastic potential energy variations correspondingly and finally $\bar k\to
{\bm K}={\bm A}\approx{\bm a}/c^2$,  then we get from (\ref{laplace}) ${\bm
f}=m{\bm a}$ under $q=1/2$.

Another useful analogy, supporting our interpreting of mechanics is standard
string oscillation problem. The oscillation equation: $\rho u_{tt}=T_{\rm
sp}u_{xx}$, where $T_{\rm sp}$ ---  space-like tension takes fully symmetric
form: $T_{\rm t}u_{\tau\tau}=T_{\rm sp}u_{xx}$, if one change $t\to
\sqrt{2}ct=\tau$, $2\rho c^2=T_{\rm t}$.
\item
Boundary conditions at the space-like ends of the string are unavailable for
observer, since they lie in causally non-connected with him region of
$\mathbb{M}_4$. The Saint-Venant  principle generalization, used in above
deducing  bend energy, in ordinary language means, that information about the
ends state is available only for times
 $\tau\sim\mathcal{H}/c$ near beginning and end of the string.  It is the time, that is required for  disturbance propagation
 from the ends along light corners from the  past and future ends inside 4-string.
For a 3D observer, which could be able to reach this ends regions\footnote{It
may not necessary to be an absolute beginning  or end of history of 3D body,
but also those moments of time, when the 3D body is essentially influenced by
another ones, for example the moment  of destroying   under non-elastic
collision.}, they would look like some involved and anomalous evolution of
3-body, including its non-stationary rotation and space-like deformation.
Inversely, at regions of the string remoted from its ends, 3-body has a regime
of "simple motion", which is described by classical Newtons laws.
\item
Continuous  evolution of 3-bodies in observable physical world means, that
there are no isolated 4-strings: each of them  has a beginning of its own
absolute history and its end (see Fig.\ref{fb4}), where corresponding 3-body is
naturally formed (or artificially made or destroyed) from another mother
3-bodies or splits into another daughter 3-bodies. These last  form in 4D world
its own 4-strings too, which have force (non-causal  from the SR viewpoint)
interaction with middle string  (or strings) through the space-like ends. So,
the best object, which represents the 4D mechanical structure of our Universe
from the viewpoint of present approach, is {\it space-time tense net}. 3-body
Universe of classical mechanics is its simultaneousity section. Similarly,
bounded configurations of 3-bodies (for instance planetary system or galaxy
clusters) form in Minkowski space-time {\it 4D tense ropes.}
\item
From expressions (\ref{equiv}) and  (\ref{class}) it is easy to show, that
classical action (more precisely  $c\mathfrak{s}$  in ordinary units) within
elasticity theory has the sense of  {\it 4D free strain energy}, that clears
its importance in variational formulating of classical mechanics and field
theory. We see, that Lagrange function within the former appears as its linear
1-density, and lagrangian  within the latter --- as a volume 4-density.
Standard variational mechanical problem  with fixed initial and final points of
trajectory corresponds to a {\it 4-string with supported ends}, that is one of
variety of boundary conditions, appearing within bar deformation theory. The
question about true boundary condition of 4-bars and 4-plates should be
answered by experimental way (see \cite{kok4}).
\item
Theoretical generalization of classical mechanics within  the frame of the
 approach is possible in two directions. The first one
takes into account relativistic corrections  ( for example expressions
(\ref{torfren}) in \S\ref{torsion}) for  elastic energy of twist and bending
and then compares its influence on shape of the 4-string (in  $\mathbb{V}_3^t$
--- motion law) with experiment. Second one includes non-linear over strain
tensor addenda into free energy, which concern with invalidity of Hooks law,
and fixes values of new elastic constants by experiment. Note, that in this
non-linear elasticity theory superposition principle, mentioned in the
beginning of   \S\ref{bend}, is, generally speaking, violated.
\item
The role of an observer in the approach demands the particular discussion. As a
matter of fact,  the whole dynamics (i.e. variability of the 3D world)
retranslates here into a motion  {\it of perception of an observer from one
simultaneousity hypersurface to another.} Such motion has already been
postulated in various  multidimensional theories (\cite{pav0}-\cite{pav3},
\cite{pav4} and ref-s therein). The essence  of the postulate in our approach
is that {\it without observer and its perception motion, there is no 3D
dynamics --- relative history of a mechanical system of 3-bodies but only its
absolute one.} So, the concept of observer is necessary not only within the
frame of quantum mechanics for resolution of some paradoxes \cite{kv,pav4}, but
also in classical physics for formulating of observable laws of physical world.
\end{enumerate}

We believe, that our approach can be useful, at least, from the two points of
view. On the one hand, it clears  simple and, at the same time, deep 4D origin
of laws of classical mechanics and gives obvious ways of its generalization. On
the other hand, it also clears general origin of kinetic  parts  of our
lagrangians, while more often physicists  hint on  potential ones. Using the 4D
elastic representation, one can say, that {\it there are no proper kinetic
terms in classical mechanics: they also have a sense of twist and bend
potential energies of 4-string, when 4D Hooks law takes place.} Note, that
4-energy density $\Upsilon$ in (\ref{pot4}) can be also treated in term of 4D
elasticity theory: we only should consider 4-strings as {\it linear stresses at
4D plate}, which  interact with each other by short range forces, described by
stress tensor at the plate. In this picture the plate (or space-time itself)
appears as elastic medium with interesting and unusual properties
(\cite{kok1,kok2,kok3}).

\vspace{1cm} {\centering\large\bf ACKNOWLEDGEMENTS}

\vspace{0.5cm} I would like to express many thanks to prof. M.Pav\u{s}i\u{c},
prof. Yu. S. Vladimirov, prof. V. G. Krechet, dr. V. A. Korotkiy, dr. A. V.
Soloviev for useful critical comments, to my friend Kyrill Goodz for assisting
in translation and to E. P. Stern for technical  supporting and also to all
participants of seminar RGS "Geometry and physics" for useful discussions.

\onecolumn Table 1. Assumptions, revealing the properties of 4-D elastic thin
bars relevant to observable ordinary 3D rigid bodies in classical mechanics.
\vspace{1cm}

\begin{tabular}{|c|l|l|}
\hline $N$ & 4-elasticity language& mechanical language\\ \hline 1.&Flat
sections hypothesis $({\bm\psi}=0)$& Absolutely rigid 3-body\\ \hline 2.&Weak
twist $({\bm \omega}\mathcal{H}\ll1)$& Nonrelativistic rotation\\ \hline
3.&$\zeta=\rho c^2$ & Elastic  nature of rotational inertia\\ \hline 4.&Bar is
tense string ($F_{b}\ll F_{s}$)& second order of Newton equations\\ \hline
5.&Weak bending (extrastretching) ($|{\bm{} v}|\ll1$)& nonrelativistic
progressive motion\\ \hline 6.&$\rho c^2=\Pi/2$ & $t$-Force nature of mass\\
\hline
\end{tabular}

\end{document}